\documentstyle[preprint,aps]{revtex}
\date{14 November 1995}
\begin{document}
\draft
\title{Dirac Fermions with Disorder in Two Dimensions: Exact Results}
\author{K. Ziegler}
\address{Max-Planck-Institut f\"ur Physik Komplexer Systeme,
Au\ss enstelle Stuttgart, Postfach 800665, D-70506 Stuttgart, Germany}
\maketitle 
\begin{abstract}
Dirac fermions on a two-dimensional lattice with disorder are considered. The
Dirac mass, which controls the gap between the two bands of the fermions, is
subject to random fluctuations. Another type of disorder is discussed presented by a
random vector potential. It is shown that the imaginary part of the
one-particle Green's
function can be written as the imaginary part of another Green's function
which has only poles on the lower half-plane.
Therefore, it is possible to perform a Cauchy integration for
a Lorentzian distribution in analogy with the Lloyd model.
The results are compared with calculations performed in the
continuum limit based on renormalization group and bosonization methods.
\end{abstract}
\section{Introduction}
The density of states (DOS) in two-dimensional electron systems with a
pseudo-gap is a subject of interest for a number of physical situations
discussed recently \cite{fradkin,lee,hats,osh,fifra,lud,zie0,tsvel,xiang}.
A typical model with a pseudo-gap is represented by the Dirac Hamiltonian in
two dimensions
\begin{equation}
H=i\nabla_1\sigma_1+i\nabla_2\sigma_2+m\sigma_3,
\end{equation}
where $\sigma_\mu$ are Pauli matrices including the $2\times2$ unit matrix
$\sigma_0$. The Dirac equation for a state $\psi$ then is
$-\partial\psi/\partial t=H\psi$.
The dispersion relation is $E(k_1,k_2)=\pm\sqrt{m^2+k_1^2+k_2^2}$ in the
continuum limit. (The lattice will be considered later.) The two signs
describe the particle and the hole band, respectively. Both bands touch each
other if the Dirac mass vanishes, as one can see in the DOS $\rho(E)\propto
|E|\Theta(E^2-m^2)$, where $\Theta$ is the step function. The touching bands is
also a feature of a second order phase transition
because the decay length of the corresponding Green's function diverges as one
goes to the special (critical) point $m=0$. This behavior is indeed formally
related to a number of critical phenomena in two-dimensional systems like the
ferromagnetic phase transition of the two-dimensional Ising model 
\cite{dotsenko}.
Another physical example, described by the Dirac Hamiltonian, is the
degenerate semiconductor which exists for $m=0$ \cite{fradkin}. Furthermore,
the large scale limit of a two-dimensional electron gas on a square lattice
near the integer quantum Hall transition for certain commensurate flux
situations (e.g., half a flux quantum per plaquette) 
is described by Dirac fermions \cite{hats,osh,fifra,lud,zie0}.
A common
feature of all these systems is that the DOS at the touching bands
(i.e., at $m=0$) is zero, i.e., there is a pseudo-gap. This raises the
question whether there is a mechanism which creates states in the
pseudo-gap,
for instance, electron-electron interaction or quenched disorder.
This is important in order to understand if there is a non-vanishing
density of low-energy excitations created by interaction or disorder.
In this article only the effect of quenched disorder will be analyzed.

There is a number of studies for the effect of disorder in the pseudo-gap
of Dirac fermions.
A numerical calculation for an electron on a square lattice with
half a flux quantum per plaquette shows a non-zero density at low
energies \cite{hats}. 
A coherent potential approximation (CPA) of the Dirac fermions with random
energy term $E\sigma_0$ added to $H$ also gives a non-zero DOS \cite{fradkin}.
A similar result was found for a random mass term $m\sigma_3$ in
a modified model with $N$ fermion levels per site, using the $N\to\infty$
limit \cite{zie00}. However, these are essentially mean-field results
which may be affected strongly by fluctuations in the two-dimensional
system. It is possible that the CPA or $N\to\infty$ result are destroyed
by fluctuations
in $d=2$. Therefore, as an alternative approach a renormalization group
treatment was applied to this problem \cite{dotsenko}. From this it turned out
that there is asymptotic freedom indicating that the pseudo-gap, which is
controlled by large scale degrees of freedom, is not affected by a random
Dirac mass (marginally irrelevant perturbation).
However, a rigorous estimation leads to a non-zero lower bound of the DOS in
the pseudo-gap \cite{zie1}, at least for a random Dirac mass. The
renormalization group calculation indicates that the random energy term
is a relevant perturbation, in agreement with the CPA result. A third type
of disorder was studied recently by adding a random vector potential to $H$
\cite{lud}. The renormalization group and bosonization treatment indicate
a more complicated behavior of the pseudo-gap in this case:
The average DOS vanishes like $|E|^\alpha$ with a non-universal
exponent $\alpha>0$ if the randomness is weaker than a critical strength.
On the other hand, the average DOS diverges if the randomness is stronger
than the critical strength because of $\alpha<0$. (For more detailed
results see Sect. III.)

A similar system with a pseudo-gap is the d-wave superconductor.
Nersesyan et al. \cite{tsvel}
analyzed this system in $d=2$ and found for the pseudo-gap of the average DOS
$\rho(E)\sim |E|^{1/7}$ in contrast to the linear behavior of the pure system.   
However, this result is in disagreement with others which also find
a destruction of the pseudo-gap  \cite{xiang}. The effect of disorder in the
d-wave superconductor will be discussed in a separate article.
The aim of this article is to present an exact solution for the imaginary part
of the single particle Green's function of disordered Dirac fermions.

\noindent
There are several examples in the theory of a quantum particle in a random
potential where the average one-particle Green's function can be calculated
exactly.
Apart from a number of one-dimensional examples \cite{LGP}, there is the
Lloyd model \cite{Lloyd}. It is defined by the Hamiltonian $H=H_0+V$, where
$H_0$ is a Hermitean matrix
(e.g., a tight-binding Hamiltonian for a particles on a lattice). $V$ is a
random potential distributed according to a Lorentzian (or Cauchy) distribution
\begin{equation}
P(V)dV=(\tau/\pi)[\tau^2+(V-V_0)^2]^{-1}dV
\label{1}
\end{equation}
The distribution density has two poles: $V=V_0\pm i\tau$.
The Green's function $G(z)=(H+z)^{-1}$ must be averaged with respect to the
random potential. $G(z)$ as a function of $V_x$ for a fixed site $x$ is
analytic
in the upper (lower) complex half-plane if $\Im z$ is positive (negative),
respectively. Therefore, the path of integration of $V_x$ can be closed in that
half-plane where $G$ is analytic. As a result, only the pole of the Lorentzian
density contributes to the integral $\int G P(V_x)dV_x$.
This integration can be performed for all lattice sites
leading eventually to the average Green's function
\begin{equation}\langle G(z)\rangle =[H_0+V_0+z+i sign(\Im z)\tau]^{-1}
\equiv G(z+i sign(\Im z)\tau).
\label{2}
\end{equation}
The average DOS then reads
\begin{equation}\langle\rho(E)\rangle =-(1/N\pi)\lim_{\epsilon\to0}\Im Tr
\langle G(E+i\epsilon)\rangle
,\label{3}
\end{equation}
where $Tr$ is the trace operator and $N$ is the number of lattice sites.
Another examples for an exact solution
is the DOS of a particle in a homogeneous magnetic field in
two dimensions. If the corresponding Hilbert space of the particle is projected
onto the lowest Landau level \cite{Weg}, the average DOS for a white noise
potential can be calculated exactly by summing up all terms of the
perturbation theory with respect to the white noise potential.
The exact solution is related to the fact that the lowest Landau level system
is equivalent to a zero-dimensional model. 
It was discovered by Br\'ezin et al. \cite{BGI} that the latter is a
manifestation of the dimensional reduction of the two-dimensional system by 2
due to a supersymmetry of the lowest Landau level problem.
Unfortunately, the simplicity of the average DOS of the lowest Landau level
cannot be extended to higher Landau levels. It
is also in sharp contrast to the complexity of the description of the
localization properties \cite{Pruis}.

\noindent
There is some hope that the treatment of an electron on the square lattice in
a strong magnetic field is simpler than a continuum model. The lattice model
is motivated by numerical simulations \cite{chalk} and analytic
calculations \cite{osh,fifra,lud,zie0}. The reason for a simplification is
that the electron near a quantum Hall transition behaves like a Dirac fermion
\cite{osh,fifra} because the excitations near the Fermi energy have a linear
dispersion. The Hamiltonian of the Dirac fermions on a square lattice with
unit lattice constant is
\begin{equation}
H+i\epsilon\sigma_0=(i\nabla_1+a)\sigma_1+i\nabla_2\sigma_2+m\sigma_3
+i\epsilon\sigma_0.
\label{4}
\end{equation}
The lattice gradient
$i\nabla_\mu$, with $\nabla_\mu f(x)=(1/2)[f(x+e_\mu)-f(x-e_\mu)]$ and lattice
unit vectors $e_1$, $e_2$, is Hermitean.
Two types of disorder are discussed subsequently: a random Dirac mass $m$
and a random vector potential $a$. The vector potential term is chosen in
(\ref{4}) in the same way as in Ref.\cite{lud}. It can be considered as a weak
disorder approximation of a fluctuating Peierls phase factor in Landau gauge.
The Green's function now reads $G(m,i\epsilon)=(H+i\epsilon\sigma_0)^{-1}$,
i.e., $m$ and $\epsilon$ correspond
with the real and imaginary part of $z$ in the Green's function of the Lloyd
model, respectively. The treatment of this problem is rather technical,
although the aim is always to find a $G$ with $\Im G=\Im G'$, where the
analytic properties of $G$ and $G'$ are different:
$G=(H+i\epsilon\sigma_0)^{-1}$, as a function of a random variable at a given
site, has poles on {\it both} complex half-planes whereas
$G'=(H'+i\epsilon\sigma_0)^{-1}$ has only a pole on {\it one} of the complex
half-plane. The Hamiltonian $H'$ is obtained from $H$ by multiplication with
a diagonal matrix. The latter depends on the specific type of randomness.

The article is organized as follows. In Sect. II the random Dirac mass and in
Sect. III a random vector potential are analyzed. The problem of species
multiplication due to the lattice is discussed in Sect. IV and the
projection onto the homogeneous modes on the lattice is given. 

\section{Random Dirac Mass}

The matrix $H+i\epsilon\sigma_0$
depends on the two complex variables $\pm m_x+i\epsilon$. Thus, in contrast to
the Lloyd model, $G(m,i\epsilon)$ may have singularities in both complex
half-planes. Therefore,
the Green's function of Dirac fermions is similar to the two-particle
Green's function of a non-relativistic particle. However, it will be shown
subsequently that there is an alternative representation for the imaginary
part of the Green's function which depends only on a single complex variable
like the Green's function of the Lloyd model. As a first step, 
$H+i\epsilon\sigma_0$ is multiplied by a diagonal matrix $D\sigma_3$ from the
right ($D$ is the staggered diagonal matrix $D_{x,x'}=(-1)^{x_1+x_2}
\delta_{x,x'}$ with the two-dimensional space coordinates $x=(x_1,x_2)$)
\begin{equation}
H'=i(i\nabla_1)D\sigma_2+i(i\nabla_2)D\sigma_1+mD\sigma_0+i\epsilon D\sigma_3,
\label{5}\end{equation}
where $\nabla_\mu D$ is Hermitean, since $D$ anticommutes with $\nabla_\mu$.
Hermitean conjugation of $H'$ yields
\begin{equation}{H'}^\dagger=i(i\nabla_1)D\sigma_2+i(i\nabla_2)D\sigma_1+mD
\sigma_0-i\epsilon D\sigma_3.
\label{6}
\end{equation}
Moreover, $\sigma_3$ anticommutes
with $\sigma_1$ and $\sigma_2$. Consequently, $i\epsilon D\sigma_3$ {\it
commutes} with all other terms in $H'$. These properties lead to the product
\begin{equation}H'{H'}^\dagger=[i(i\nabla_1)D\sigma_2+i(i\nabla_2)D\sigma_1+mD
\sigma_0]^2+\epsilon^2\sigma_0.
\label{7}
\end{equation}
From the definition of $H'$ follows directly
\begin{equation}H'{H'}^\dagger=
[H+i\epsilon\sigma_0]\sigma_3D\sigma_3D[H+i\epsilon\sigma_0]^\dagger
=[H+i\epsilon\sigma_0][H-i\epsilon\sigma_0]= [H-i\epsilon\sigma_0][H+
i\epsilon\sigma_0].
\label{8}
\end{equation}
The r.h.s. of (\ref{7}) can also be written $H''{H''}^\dagger$ with
\begin{equation}
H''=i(i\nabla_1)D\sigma_2+i(i\nabla_2)D\sigma_1+(mD+i\epsilon)\sigma_0.
\label{9}
\end{equation}
As a result, $H''$ depends only on one complex variable $(-1)^{x_1+x_2}m_x
+i\epsilon$ for a given site $x$.
The imaginary part of the Green's function $(H+i\epsilon\sigma_0)^{-1}$
reads
$(i/2)([H+i\epsilon\sigma_0]^{-1}-[H-i\epsilon\sigma_0]^{-1})=
\epsilon({[H-i\epsilon\sigma_0][H+i\epsilon\sigma_0]})^{-1}$,
i.e., it depends on the Hamiltonian only via $H^2$. Therefore, the identity
$H^2+i\epsilon^2\sigma_0=H''{H''}^\dagger$ can be used to write
\begin{eqnarray}
{i\over2}([H+i\epsilon\sigma_0]^{-1}-[H-i\epsilon\sigma_0]^{-1})
=\epsilon({[H-i\epsilon\sigma_0][H+i\epsilon\sigma_0]})^{-1}
=\epsilon(H''^\dagger H'')^{-1}
\nonumber\\
={i\over2}[H''^{-1}-(H''^\dagger)^{-1}].
\label{main}
\end{eqnarray}
Thus the imaginary part of the average Green's function
$(H-i\epsilon\sigma_0)^{-1}$ can be
calculated exactly for a Lorentzian distribution due to (\ref{main}),
where $m_x$ can be integrated out explicitly as in the Lloyd model. Only a
pole of the distribution contributes leading to the replacements
$\epsilon\to\tau+\epsilon$ and $m\to m_0$ in $H$. This implies
\begin{equation}
{i\over2}\langle[H+i\epsilon\sigma_0]^{-1}-[H-i\epsilon
\sigma_0]^{-1})\rangle
={i\over2}[{\bar H}^{-1}-({\bar H}^\dagger)^{-1}]
\label{maina}
\end{equation}
with
\begin{equation}
{\bar H}=i\nabla_1\sigma_1+i\nabla_2\sigma_2+i(\epsilon+\tau)\sigma_0+m_0
\sigma_3.
\label{13}
\end{equation}
The imaginary $\tau$-term leads always to an exponential
decay of the average Green's function with a typical decay
length $\xi\sim (m_0^2+\tau^2)^{-1/2}$ at $\epsilon=0$.
Moreover, from equ. (\ref{3}) follows for the average DOS
\begin{equation}
\langle\rho(i\epsilon,m_0)\rangle=-{1\over N\pi}\Im Tr[{\bar 
H}^{-1}].
\label{12}\end{equation}
The dependence on the energy $E$ is obtained from an analytic continuation
$i\epsilon\to i\epsilon+E$.
The resulting average DOS is plotted in Fig.1 for $\tau=0.01$ and
in Fig.2 for $\tau=0.1$. The non-vanishing DOS is in agreement with a
rigorous
proof \cite{zie1} and a numerical result \cite{hats}. For Gaussian disorder
with variance $g$ there is a lower bound \cite{zie1}
\begin{equation}
\langle\rho(0,0)\rangle\ge c_1e^{-c_2/g}.
\end{equation}
with some positive constants $c_1$, $c_2$, independent of $g$.

In the continuum limit it was argued, using a one-loop renormalization group
calculation, that random fluctuations of the Dirac mass are irrelevant on
large scales \cite{lud,tsvel}. This implies a linearly vanishing DOS and a
divergent correlation length at $E=m_0=0$.

\section{Random Vector Potential}

A calculation analogous to that of the random mass system can be performed for
a random vector potential. For this purpose the orthogonal transformation
$(\sigma_1+\sigma_3)/\sqrt{2}$ is
applied to the Hamiltonian
\begin{equation}
H+i\epsilon\sigma_0\to(i\nabla_1+a)\sigma_3-i\nabla_2\sigma_2+m\sigma_1
+i\epsilon\sigma_0.
\label{20}
\end{equation}
The multiplication of the massless Hamiltonian (i.e., $m=0$) from the r.h.s.
with $D'\sigma_3$ (where $D'_{x,x'}=(-1)^{x_2}\delta_{x,x'}$) yields 
\begin{equation}
H'=(i\nabla_1+a)D'\sigma_0-i(i\nabla_2)D'\sigma_1+i\epsilon D'\sigma_3.
\label{21}
\end{equation}
The lattice difference operators
$i\nabla_1D'$ and $i(i\nabla_2)D'$ are Hermitean. Since $D'\sigma_3$ commutes
with the first two terms of $H'$, one obtains
\begin{equation}
[H+i\epsilon\sigma_0][H-i\epsilon\sigma_0]=H'H'^\dagger=[(i\nabla_1+a)D'
\sigma_0-i(i\nabla_2)D'\sigma_1]^2+
\epsilon^2\sigma_0=H''{H''}^\dagger
\label{22}
\end{equation}
with
\begin{equation}
H''=(i\nabla_1+a)D'\sigma_0-i(i\nabla_2)D'\sigma_1++i\epsilon\sigma_0
\label{23}
\end{equation}
which can be used to establish again equation (\ref{main}).
For $\langle a\rangle=0$ the average imaginary part of the Green's function
and, therefore, the average DOS for $m_0=0$ is related to the Hamiltonian
${\bar H}$ as given in (\ref{13}).

In contrast to this result, the bosonization of the Dirac fermions
in the continuum limit leads to a different behavior \cite{lud}.
For instance, the DOS reads
\begin{equation}
\langle\rho(E,0)\rangle\sim E^{(2-z)/z},
\end{equation}
where $z=1+\Delta_A/\pi$ ($\Delta_A$ is the variance of the fluctuations
of the vector potential). I.e., the DOS vanishes at $E=0$ for $z<2$ (weak
disorder) and diverges for $z>2$ (strong disorder). The Green's function
behaves like
\begin{equation}
\langle G_{0,x}(E,m=0)\rangle\sim e^{i|x|/\lambda}e^{-|x|/\xi_1}
\end{equation}
where for $E\sim0$
\begin{equation}
\lambda\sim E^{-(1-\Delta_A/\pi)}
\end{equation}
\begin{equation}
\xi_1\sim E^{-1}/\Delta_A.
\end{equation}
Thus the Green's function decays exponentially for $E\ne0$. There is
a critical point $E=0$
where the correlation length diverges with $E^{-1}$.
The difference of the results of \cite{lud} and the present work
is probably related to the order of taking the
continuum limit and averaging over disorder. It is not a consequence of the
difference
of disorder distributions (Gaussian in \cite{lud} versus Lorentzian
distribution here) because the Gaussian distribution could also be treated for
${H''}^{-1}-{H''^\dagger}^{-1}$ in a strong disorder expansion.
The result of this expansion is also a finite non-zero DOS and a finite
correlation length of $G$.
\section{Remark on Species Multiplication}
The phenomenon of species multiplication in a fermion lattice theory
is well-known from lattice gauge field theories \cite{kogut}. It is due to
several nodes in the energy dispersion of the lattice
model which indicate the existence of low energy excitations on different
length scales.
The dispersion of the Dirac fermions considered in this article for $m=a=0$
is $E(k_1,k_2)=\pm\sqrt{\sin^2k_1+\sin^2k_2}$. It has 9 nodes at 
$k_j=0,\pm\pi$ (cf. Fig.3).
In contrast to the lattice model the corresponding continuum
model, with $E(k_1,k_2)=\pm\sqrt{k_1^2+k_2^2}$,
has low energy excitations only for small wavevectors (i.e., on large
scales) as discussed in the Introduction. It will be shown in this Section,
using the random mass model of Section II, that the species multiplication
is not the reason for the smooth properties of the one-particle Green's
function.

The degeneracy of the low energy behavior of the lattice model can be
lifted by introducing additional terms in the Hamiltonian \cite{kogut}.
A possible way is to replace the Hamiltonian $H$ by the
new Hamiltonian $H+\delta(\Delta-2)\sigma_3$, where $\delta$ is
a positive number ($0<\delta\le1$) and 
$\Delta$ is a lattice operator with $\Delta f(x)= [f(x+e_1)+f(x-e_1)+
f(x+e_2)+f(x-e_2)]/2$. The dispersion of the new $H$ is 
$E(k_1,k_2)=\pm\sqrt{\delta^2(\cos k_1+\cos k_2-2)^2+\sin^2k_1+\sin^2k_2}$
for $m=a=0$. This is shown in Fig.4 for $\delta=1/2$. 
It is not clear to the author which tranformation can be
applied to relate the imaginary part Green's function with the new Hamiltonian
in order to get the analytic behavior necessary to perform the Cauchy
integration with respect to the randomness. However, this difficulty can be
circumvented by generalizing $H$ to ${\hat H}$ with
\begin{equation}
{\hat H}=\pmatrix{
H+\delta(\Delta-2)\sigma_3&m'\sigma_3\cr
m'\sigma_3&H-\delta(\Delta+2)\sigma_3\cr
},
\end{equation}
where $m'$ is a random variable which is statistically independent of $m$ with
mean zero.
Now the orthogonal transformation
\begin{equation}
{1\over\sqrt{2}}\pmatrix{
\sigma_0&\sigma_0\cr
\sigma_0&-\sigma_0\cr
}
\end{equation}
rotates the diagonal part $(\Delta\sigma_3,-\Delta\sigma_3)$ in the
off-diagonal positions and the off-diagonal part into the diagonal position
$(m'\sigma_3,-m'\sigma_3)$ such that
\begin{equation}
{\hat H}=\pmatrix{
H-(2\delta-m')\sigma_3&\Delta\sigma_3\cr
\Delta\sigma_3&H-(2\delta+m')\sigma_3\cr
}.
\end{equation}
The random variables $M_x\equiv -2\delta+m_x+m_x'$ and $
M'_x\equiv -2\delta+m_x-m_x'$ in the diagonal part
of ${\hat H}$ can now be considered as new independent random variables.

The transformation
\begin{eqnarray}
{\hat H}\to\pmatrix{
\sigma_0&0\cr
0&-\sigma_0\cr
}{\hat H}\pmatrix{
D\sigma_3&0\cr
0&D\sigma_3\cr
}
\nonumber\\
=\pmatrix{
HD\sigma_3-(2\delta-m')D\sigma_0&\Delta D\sigma_0\cr
D\Delta\sigma_0&-HD\sigma_3-(2\delta+m')D\sigma_0\cr
}={\hat H}'
\label{trans}
\end{eqnarray}
generates the Hermitean matrix ${\hat H}'$.

Using for the r.h.s. of (\ref{trans}) the notation $T_0{\hat H}T_1$ and
applying the property
\begin{equation}
T_0{\hat H}T_1=T_1{\hat H}T_0
\label{comm}
\end{equation}
one obtains
\begin{equation}
({\hat H}'-i\epsilon D\gamma_3)({\hat H}'+i\epsilon D\gamma_3)=
T_0({\hat H}-i\epsilon\gamma_0)T_1T_0({\hat H}+i\epsilon\gamma_0)T_1
=T_0({\hat H}-i\epsilon\gamma_0)({\hat H}+i\epsilon\gamma_0)T_0
\label{prod1}
\end{equation}
with the diagonal matrices $\gamma_0=(\sigma_0,\sigma_0)$ and
$\gamma_3=(\sigma_3,-\sigma_3)$.
Moreover, one has for the imaginary part of the one-particle Green's
function as before
$(i/2)\Big[({\hat H}+i\epsilon\gamma_0)^{-1}
-({\hat H}-i\epsilon\gamma_0)^{-1}\Big]
=\epsilon\Big[({\hat H}-i\epsilon\gamma_0)({\hat H}
+i\epsilon\gamma_0)\Big]^{-1}$.
Due to (\ref{prod1}) and $T_0^{-1}=T_0$ this can be rewritten as
\begin{equation}
\epsilon
T_0\Big[({\hat H}'-i\epsilon D\gamma_3)({\hat H}'+i\epsilon D\gamma_3)\Big]
^{-1}T_0.
\label{prod2}
\end{equation}
The l.h.s. of (\ref{prod1}) reads
\begin{equation}
({\hat H}'-i\epsilon D\gamma_3)({\hat H}'+i\epsilon D\gamma_3)=
({\hat H}')^2+\epsilon^2\gamma_0=
({\hat H}'-i\epsilon\gamma_0)({\hat H}'+i\epsilon\gamma_0)
\end{equation}
because $H'$ and $D\gamma_3$ commute. This implies for (\ref{prod2})
\begin{equation}
\epsilon
T_0\Big[({\hat H}'-i\epsilon\gamma_0)({\hat H}'+i\epsilon\gamma_0)\Big]
^{-1}T_0
={i\over2}\Big[({\hat H}'+i\epsilon\gamma_0)^{-1}
-({\hat H}'-i\epsilon\gamma_0)^{-1}\Big].
\end{equation}
Consequently, the imaginary part of the Green's function satisfies
\begin{equation}
{i\over2}\Big[({\hat H}+i\epsilon\gamma_0)^{-1}
-({\hat H}-i\epsilon\gamma_0)^{-1}\Big]
={i\over2}T_0\Big[({\hat H}'+i\epsilon\gamma_0)^{-1}
-({\hat H}'-i\epsilon\gamma_0)^{-1}\Big]T_0,
\end{equation}
analogously to (\ref{main}).
At a given site $x$ the matrix ${\hat H}'+i\epsilon\gamma_0$ depends on 
the random variables in the combinations $(-1)^{x_1+x_2}M_x+i\epsilon$ and
$(-1)^{x_1+x_2}M'_x+i\epsilon$. Assuming a Lorentzian distribution for $M_x$
and $M'_x$, the integration can be performed again as in Sect. II. As a
result the imaginary part of the average Green's function is
\begin{equation}
\Im\pmatrix{
{\bar H}+\delta(\Delta-2)\sigma_3&0\cr
0&{\bar H}-\delta(\Delta+2)\sigma_3\cr
}^{-1},
\end{equation}
where ${\bar H}$ is the average Hamiltonian (\ref{13}).
Thus the lifting of the degeneracy of the nodes in the dispersion relation
does not change the analytic behavior of the average one-particle Green's
function.

\section{Conclusion}
An exact expression for the average imaginary part of the one-particle
Green's function and the average DOS of two-dimensional lattice
Dirac fermions have been derived for a random Dirac mass and for a random
vector potential. We have shown that there is a non-zero DOS
due to disorder and there is a finite decay length for the
average one-particle Green's function. This implies the creation of a
non-vanishing density of low-energy excitations due to disorder in a vicinity
of $E=M=0$.
These lattice results are in agreement with numerical simulation \cite{hats}.
However, they are in disagreement with the results of
a renormalization group calculation and a bosonization approach for a
continuous system of Dirac fermions \cite{lud,tsvel}, where the DOS vanishes
or diverges at $E=M=0$. Moreover, the lattice model does not exhibit the
critical properties of the Green's function and the DOS found in the
renormalization group calculation and in the bosonization approach. It is
possible to take the continuum limit of the lattice model after the averaging
over disorder, for instance, in the Hamiltonian (\ref{13}). This, however, does
not lead to a critical behavior. It seems that the critical behavior of the DOS
is a consequence of taking the continuum limit first and performing the
averaging over disorder afterwards. This is plausible because the effect of
randomness is much stronger in the continuum due to statistically independent
fluctuations on arbitrarily short scales. 
It is shown in Sect. IV that species multiplication, which is a special effect
of the lattice model, is not the reason for the smooth behavior of the average
DOS.

\noindent
Acknowledgement: I am grateful to D. Braak for interesting discussions.

\begin{figure}
\caption{Average density of states for disorder strength $\tau=0.01$.}
\caption{Average density of states for disorder strength $\tau=0.1$.}
\caption{Energy dispersion $E(k_1,k_2)$ of lattice Dirac fermions.}
\caption{Energy dispersion of lattice Dirac fermions after lifting
         the degeneracy of the low energy properties}
\end{figure}

\begin{references}
\bibitem{fradkin}
E.Fradkin, Phys.Rev. {\bf B33}, 3257 (1986), ibid. 3263 (1986)

\bibitem{lee}
P.A.Lee, Phys.Rev.Lett. {\bf 71}, 1887 (1993)

\bibitem{hats}
Y.Hatsugai and P.A.Lee, Phys.Rev. {\bf B48}, 4204 (1993)

\bibitem{osh}
Y.Hatsugai, Phys.Rev. {\bf B48}, 11851 (1993), M.Oshikawa, Phys.Rev. 
{\bf B50}, 17357 (1994)

\bibitem{fifra}
M.P.A.Fisher and E.Fradkin, Nucl.Phys. {\bf B251} [FS13], 457 (1985)

\bibitem{lud}
A.W.W.Ludwig, M.P.A.Fisher, R.Shankar and G.Grinstein, Phys. Rev. {\bf B50},
7526 (1994)

\bibitem{zie0}
K.Ziegler, Europhys.Lett. {\bf 28}, 49 (1994)

\bibitem{tsvel}
A.A.Nersesyan, A.M.Tsvelik and F.Wenger, Nucl.Phys. {\bf B438}, 561 (1995)

\bibitem{xiang}
T.Xiang and J.M.Wheatley, Phys.Rev. {\bf B51}, 11721 (1995)

\bibitem{dotsenko}
V.Dotsenko and Vl.S.Dotsenko, Adv.Phys. {\bf 32} 129 (1983)

\bibitem{zie00}
K.Ziegler, Nucl.Phys. {\bf B344}, 499 (1990)

\bibitem{zie1}
K.Ziegler, Nucl.Phys. {\bf B285} [FS19], 606 (1987)

\bibitem{LGP}
I.M.Lifshits, S.A.Gredeskul and L.A.Pastur, {\sl Introduction to the Theory of
Disordered Systems} (Wiley, New York 1988)

\bibitem{Lloyd}
P.Lloyd, J.Phys. {\bf C2}, 1717 (1969)

\bibitem{Weg}
F.Wegner, Z.Phys. {\bf B51}, 279 (1983)

\bibitem{BGI}
E.Br\'ezin, D. Gross and C. Itzykson, Nucl.Phys. {\bf B235} [FS11], 24 (1984)

\bibitem{Pruis}
A.M.M.Pruisken, in {\sl The Quantum Hall Effect}, edited by R.E.Prange and
S.M.Girvin (Springer-Verlag, New York, 1990)

\bibitem{chalk}
J.T.Chalker and P.D.Coddington, J.Phys. {\bf C21}, 2665 (1988)

\bibitem{kogut}
J.B.Kogut, Rev.Mod.Phys. {\bf 55}, 775 (1983)

\end{references}
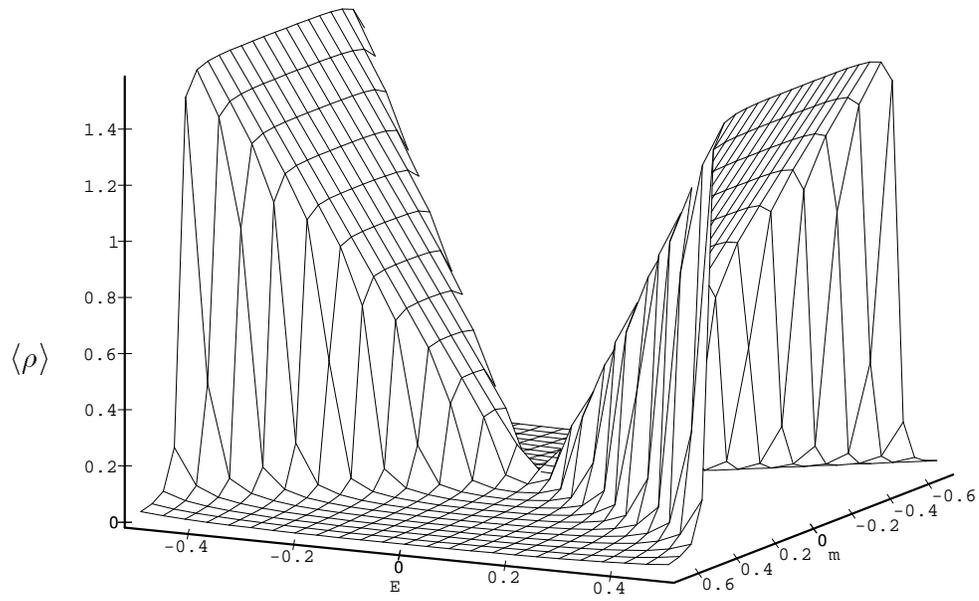
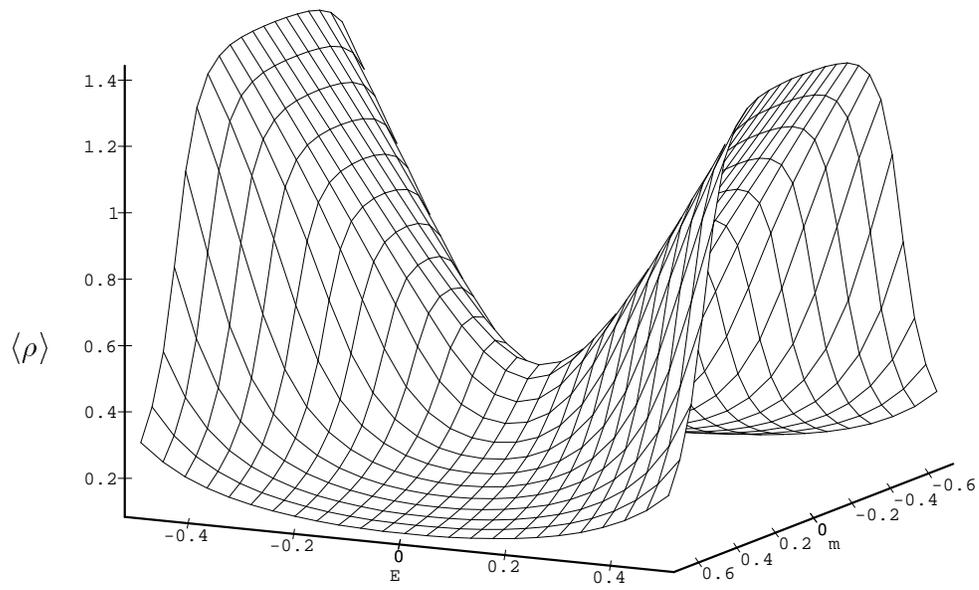
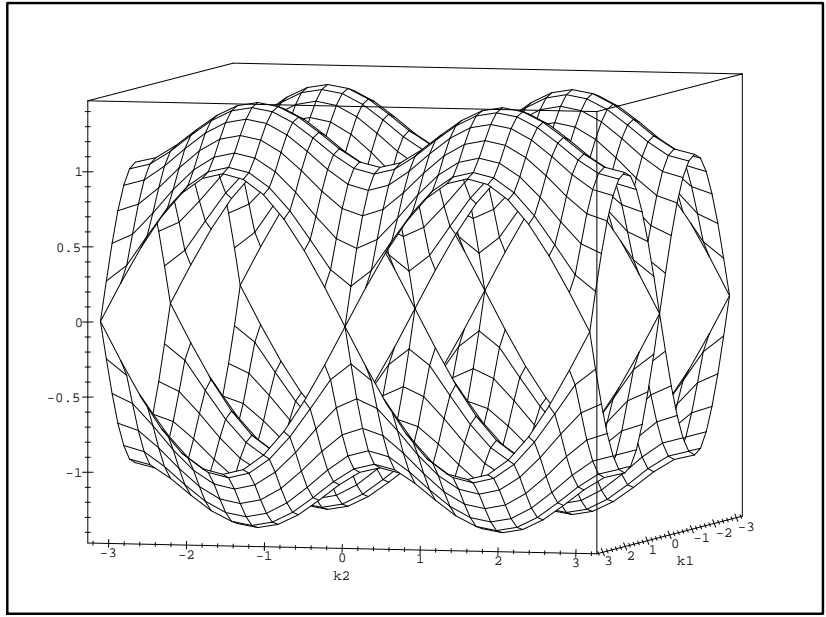
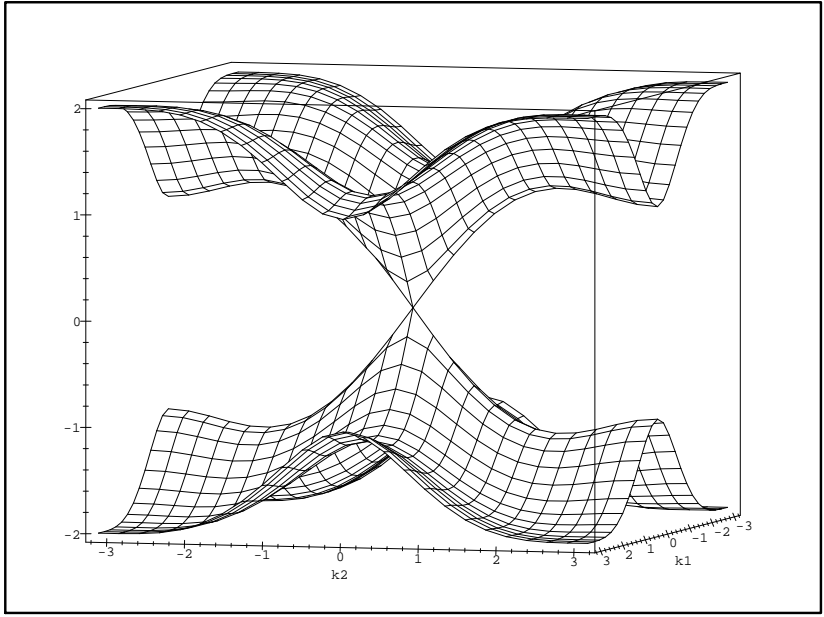
\end{document}